# CMOS-integrated superparamagnetic tunnel junction-based p-bit

Ju-Young Yoon[1*], Nuno Caçoilo[1*], Advait Madhavan[2], Jabez J. McClelland[2], Shun Kanai[1,3-7], Hideo Ohno[1,4,5,8], Shunsuke Fukami[1,3-5,8,9], William A. Borders[2]

[1]Laboratory for Nanoelectronics and Spintronics, Research Institute of Electrical Communication, Tohoku University, Sendai, Japan

[2]Physical Measurement Laboratory, National Institute of Standards and Technology, Gaithersburg, Maryland, USA

[3]Department of Electronic Engineering, Tohoku University, Sendai, Japan

[4]WPI-Advanced Institute for Materials Research (WPI-AIMR), Tohoku University, Sendai, Japan

[5]Center for Science and Innovation in Spintronics, Tohoku University, Sendai, Japan

[6]Division for the Establishment of Frontier Sciences, Tohoku University, Sendai, Japan

[7]National Institutes for Quantum Science and Technology, Takasaki, Japan

[8]Center for Innovative Integrated Electronic Systems, Tohoku University, Sendai, Japan

[9]Inamori Research Institute for Science, Kyoto, Japan

[*]These authors contributed equally

Email: s-fukami@riec.tohoku.ac.jp, william.borders@nist.gov

Probabilistic computers offer promising solutions for computationally hard problems in domains such as combinatorial optimization and machine learning. A key building block in these systems is the probabilistic bit (p-bit), which relies on superparamagnetic tunnel junctions (sMTJs) as its source of randomness. A challenging threshold to cross for scaling sMTJ-based p-bit systems is integration of sMTJs with CMOS technology. In this work, we present experimental results of a p-bit unit cell using sMTJs integrated with 130 nm CMOS technology and demonstrate that the sMTJ's resistance fluctuations can generate a corresponding fluctuating digital output voltage which is tunable via the input voltage. These findings establish the feasibility of CMOS-compatible, sMTJ-based probabilistic circuits and mark a key step toward scalable hardware for real-world probabilistic computing applications.

Computationally hard tasks like combinatorial optimization and machine learning are deemed difficult for conventional deterministic computers. Probabilistic computing is an alternative computing method using probabilistic bits (p-bits) fluctuating in time between a digital high and low state [1]. When computationally hard problems are formulated into an Ising Hamiltonian [2 these fluctuations allow the computer to probabilistically sample a multitude of configurations. Recent probabilistic computers using field-programmable gate arrays (FPGAs) have demonstrated increasingly larger problem sizes [3] by utilizing concepts like parallel-tempering [4] and sparse encoding [5]. To tackle harder problems, forecasts state that the most efficient p-computers will use unstable nanomagnets as the source of randomness [1].

A superparamagnetic tunnel junction (sMTJ) is one such unstable nanomagnet with potential for providing the randomness in a p-bit. sMTJs are composed of ferromagnetic free and reference layers separated by a tunnel barrier. The sMTJ can show a high or low resistance, depending on the angle of orientation between the two ferromagnetic layers owing to the tunneling magnetoresistance (TMR) effect [6]. In contrast to magnetic tunnel junctions (MTJs) used in magnetoresistive random access memory [7][8], sMTJs feature drastically reduced energy barriers so thermal fluctuations cause the state of the device to probabilistically change on a nanosecond timescale [9][10].

So far, sMTJ-based p-bits have been used to solve a multitude of problems [11]-[16], and are typically represented with a unit cell circuit consisting of an sMTJ in series with an n-channel MOS (NMOS) transistor whose drain output is fed into an inverter that produces binary voltage fluctuations between high and low [17].

The NMOS gate voltage is used to change the current flowing through the sMTJ which tunes the time the sMTJ spends in one state [11][12].

Current p-bit prototypes typically consist of sMTJs wire-bonded to a circuit board with an NMOS and an inverter; however, to show their full potential for practical applications, results showing operation on a large-scale ASIC are required. Therefore, a key phase of sMTJ-based p-bit development is the systematic study of back-end-of-the-line (BEOL) integration of s-MTJs with CMOS. This involves isolating each stage of the p-bit circuit to quantitatively evaluate properties suitable for both the CMOS and the sMTJ. In this work, we successfully integrate sMTJs with 130 nm CMOS technology, evaluate challenges of integration at each stage of the circuit, and demonstrate the operation of a complete p-bit unit cell.

We systematically design a CMOS-integrated p-bit circuit to verify its functionality at each maturity stage. The 5.2 mm × 3.6 mm monolithic chip containing 150 isolated sMTJs, 240 sMTJs in series with an NMOS transistor, and 150 full-stage p-bit circuits, is fabricated using a 130-nm commercial CMOS process line and a dedicated MTJ process line at Tohoku University (Fig. 1 a,b). We design the sMTJ + NMOS circuit with electrodes for the gate, source, power-supply voltage $V_{DD}$, and output connections. For the full-stage p-bit circuit, we design a cascode stage in series with the sMTJ to reduce the effect that the fluctuating resistance of the sMTJ has on the modulation of the source-drain current. The output feeds to the input of a variable threshold controller (VTC), containing two pull-up and two pull-down transistors, that collectively represent threshold voltages from 0.7 V to 1.1 V with a 100-mV step. The VTC stage output feeds into a final inverter which produces voltage fluctuations between 0 V and 1.8 V. The circuits for the sMTJ +

NMOS and full stage are completed by patterning an sMTJ between the output (or input to VTC) pads containing vias to the CMOS and $V_{DD}$. To enable a wide range of currents, the transistors are designed with channel widths of {1, 3, 9, 27} µm, producing currents from hundreds of µA to tens of mA.

To enable BEOL integration of sMTJs, the final encapsulating CMOS metal layer is left off which gives direct contact to metal and tungsten via layers below. Leaving the final metal layer off helps keep the roughness of the BEOL interface to a minimum, but carries the risk of the tungsten vias oxidizing. To prevent this, wafers were immediately coated in photoresist, where a mild removal process was used to clean the surface before material stack deposition. Before sMTJ deposition, we performed four-point probe measurements on the top and bottom electrodes to confirm ohmic contact. We then deposit the sMTJ stack above the CMOS circuits by DC and RF magnetron sputtering at room temperature of the following stack structure: Ta(5)/ PtMn(20)/ Co(2.4)/ Ru(0.9)/ CoFeB(2)/ MgO/ CoFeB(2.1)/ Ta(5)/ Ru(5)/ Ta(50) (thicknesses in nm). The sMTJs are patterned into an elliptic shape with effective diameters of {50, 60, 70, 80} nm and aspect ratios ranging from {1, 2, 3, 4} by Ar+ ion milling and reactive ion etching (Fig. 1 c,d).

To confirm electrically tunable stochastic beavior of the isolated sMTJs, the time-averaged resistance $<R>$ is aevaluated as a function of applied current in a single sMTJ (Fig. 2). Measurements were performed by connecting DC probes to top and bottom electrodes, applying a DC current, and reading voltage fluctuations on an oscilloscope. The time-averaged voltage at each current is read through the same source-measure unit supplying the current and averaged for 100 samples taken at 10 ms intervals. As the current sweeps from −80 µA to 50 µA with a 3 µA step, the sMTJ changes from spending more time in the low resistance state to spending more time in the high resistance state. With the varied diameters and aspect ratios, we measured a wide range of low resistance values ranging from 2.5 kΩ to 8 kΩ, and TMR ratios ranging from 50 % to 100 %. We also performed measurements on the random telegraph noise signal of the sMTJs, where most devices showed fluctuations with ms dwell times (not shown).

Next, we present results of sMTJs integrated in series with a single NMOS transistor to characterize a range of gate voltages required to tune the sMTJ state probabilities (Fig. 3). In the full-stage p-bit, this gate voltage will act as the input to one p-bit cell. Figure 3a shows the time-averaged output voltage $<V_\text{out}>$ of an sMTJ as the gate voltage $V_\text{gate}$ is swept from 0.7 V to 1.0 V with a 20-mV step. Each point is averaged over 20,000 points at a 10 kHz sampling rate. Measurements are performed by applying a DC voltage to both the gate of the NMOS and $V_\text{DD}$, which connects to the top electrode of the sMTJ. As the gate voltage increases/decreases, more/less current flows through the sMTJ. Accordingly, the time that the sMTJ spends in the high or low resistance state is tuned, following the result shown in Fig. 2. The output voltage $V_\text{out}$, read on an oscilloscope at the NMOS drain, is then a function of $V_\text{sMTJ}$ subtracted from $V_\text{DD}$.

For the sMTJ shown here, this results in a voltage swing at $V_\text{out}$ of 100 mV near $V_\text{gate}$ = 0.85 V. Due to a stray magnetic field [18] from the reference layer, for most devices fabricated here, the range at which the sMTJ shows resistance fluctuations is for negative currents at zero magnetic field, while it can be shifted to a positive current regime by applying a small field along the long axis opposite the stray field (Fig. 3b). As the p-bit circuit shown later can generate only positive currents, we apply this field in the subsequent experiment. Note that the stray field can be easily tuned by optimizing the stack of the reference layer.

Finally, we verify the p-bit function with a single cell of the full stage p-bit containing electrodes for $V_{DD}$, $V_{cas}$, $V_{bias}$, $V_{out}$, and ground (inset of Fig. 4). We set $V_{cas}$ to a voltage that pins the drain of the NMOS controlled by $V_{bias}$, which is the transistor used to change the sMTJ current, $V_{DD}$ to a voltage that when subtracted with the sMTJ voltage drop will result in a fluctuating input at the VTC, and $V_{OUT}$ to the input of an oscilloscope to measure the p-bit output. As $V_{bias}$ is swept from 0.5 V to 0.8 V at a 5-mV step, each $<V_{out}>$ that is time-averaged over 200,000 points transitions from 1.8 V to 0.45 V, showing a sigmoid-like behavior (Fig. 4a). The time-series response of the p-bit at three bias points along the sigmoid (Fig. 4b) demonstrates the tunability of the p-bit output with $V_{bias}$. These results indicate that the circuit with an sMTJ integrated here functions as a p-bit. We note that ideally the inverter stage should drive outputs from 1.8 V to 0 V. This issue arises not from a design mistake, but from non-idealities in the physical properties of the CMOS (Fig. 4c). We confirm this by applying 0 V to both $V_{bias}$ and $V_{cas}$ so that $V_{DD}$ drops entirely at the input of the VTC. Doing so shows a stepped increase in the inverter output starting at $V_{DD}$ = 0.8 V, which is not shown in circuit simulations.

In this work, we have experimentally demonstrated the integration of sMTJs with CMOS to produce a full-stage p-bit circuit. To fully understand the impact of sMTJ parameters on circuit performance, it will be crucial to combine both simulation and experiment. Nevertheless, these results show the potential readiness of sMTJ-based p-bits to be used in large-scale circuits for solving computationally hard problems.

**Acknowledgments**

We thank K. V. De Zoysa, F. Shibata, R. Nomura, H. Sasaki, S. Suzuki, and R. Itokazu for their technical support. This work was made possible by the NIST-led Nanotechnology Xccelerator program that distributes open-source circuit designs for integration of novel technologies on CMOS. This work was supported by JST-ASPIRE (Grant No. JPMJAP2322), MEXT Initiative to Establish Next-generation Novel Integrated Circuits Centers (X-NICS) (Grant No. JPJ011438), JSPS KAKENHI (Grant Nos. 24H02235 and 24H00039), RIEC Cooperative Research Projects and NIST.

## References

[1] K. Y. Camsari, R. Faria, B. M. Sutton, and S. Datta, Stochastic p-Bits for Invertible Logic, Phys. Rev. X, vol. 7, no. 3, 031014, Jul. 2017, doi: 10.1103/PhysRevX.7.031014.

[2] A. Lucas, Ising formulations of many NP problems, Front. Phys. vol. 2, no. 5, Feb. 2014, doi: https://doi.org/10.3389/fphy.2014.00005.

[3] N. A. Aadit, A. Grimaldi, M. Carpentieri, L. Theogarajan, J. M. Martinis, G. Finocchio, and K. Y. Camsari, Massively parallel probabilistic computing with sparse Ising machines, Nat. Electronics, vol. 5, pp. 460-468, Jun. 2022, doi: https://doi.org/10.1038/s41928-022-00774-2.

[4] C. Delacour, M. M. H. Sajeeb, J. P. Hespanha, and K. Y. Camsari, Two-dimensional parallel tempering for constrained optimization, Phys. Rev. E, vol. 112, no. 2, pp. L023301, Aug. 2025, doi: 10.1103/mr2n-qqrb.

[5] M. M. H. Sajeeb, N. A. Aadit, S Chowdhury, T. Wu, S. Smith, D. Chinmay, A. Raut, K. Y. Camsari, C. Delacour, T. Srimani, Scalable connectivity for Ising machines: Dense to sparse, Phys. Rev. Appl., vol. 24, pp. 014005, Jul. 2025, doi: https://doi.org/10.1103/kx8m-5h3h.

[6] M. Julliere, Tunneling between ferromagnetic films, Phys. Lett. A, vol. 54, pp. 225-226, Jun. 1975, doi: https://doi.org/10.1016/0375-9601(75)901747.

[7] A. D. Kent and D. C. Worledge, A new spin on magnetic memories, Nature Nanotechnology, vol. 10, no. 187, Mar. 2015, doi: 10.1038/nnano.2015.24.

[8] S. Bhatti, R. Sbiaa, A. Hirohata, H. Ohno, S. Fukami, S. N. Piramanayagam, Spintronics based random access memory: a review, Materials Today, vol. 20, pp. 530-548, Sept. 2017, doi: https://doi.org/10.1016/j.mattod.2017.07.007.

[9] K. Hayakawa, S. Kanai, T. Funatsu, J. Igarashi, B. Jinnai, W. A. Borders, H. Ohno, and S. Fukami, Nanosecond Random Telegraph Noise in InPlane Magnetic Tunnel Junctions, Phys. Rev. Lett., vol. 126, pp. 117202, Mar. 2021, doi: https://doi.org/10.1103/PhysRevLett.126.117202.

[10] C. Safranski, J. Kaiser, P. Trouilloud, P. Hashemi, G. Hu, and J. Z. Sun, Demonstration of Nanosecond Operation in Stochastic Magnetic Tunnel Junctions, Nano Letters, vol. 21, no. 5, pp. 2040-2045, Feb. 2021, doi: https://doi.org/10.1021/acs.nanolett.0c04652.

[11] W. A. Borders, A. Z. Pervaiz, K. Y. Camsari, S. Fukami, H. Ohno, and S. Datta, Integer factorization using stochastic magnetic tunnel junctions, Nature, vol. 573, pp. 390-393, Sept. 2019, doi: https://doi.org/10.1038/s41586-019-1557-9.

[12] J. Kaiser, W. A. Borders, K. Y Camsari, S. Fukami, H. Ohno, and S. Datta, Hardware-Aware In Situ Learning Based on Stochastic Magnetic Tunnel Junctions, Phys. Rev. Appl. vol. 17, pp. 014016, Jan. 2022, doi: 10.1103/PhysRevApplied.17.014016.

[13] A. Grimaldi, K. Selcuk, N. A. Aadit, K. Kobayashi, Q. Cao, S.Chowdhury, G. Finocchio, S. Kanai, H. Ohno, S. Fukami, K. Y. Camsari Experimental evaluation of simulated quantum annealing with MTJ-augmented p-bits 2022 International

Electron Devices Meeting (IEDM), San Francisco, CA, USA, 2022, pp. 1-4, doi: 10.1109/IEDM45625.2022.10019530.

[14] N. S. Singh, K. Kobayashi, Q. Cao, K. Selcuk, T. Hu, S. Niazi, N. A. Aadit, S. Kanai, H. Ohno, S. Fukami, and K. Y. Camsari, CMOS plus stochastic nanomagnets enabling heterogeneous computers for probabilistic inference and learning, Nature Communications, vol. 15, no. 2685, Nar. 2024, doi: https://doi.org/10.1038/s41467-024-466456.

[15] J. Si, S. Yang, Y. Cen, S. Chen, Y. Huang, Z. Yao, G.-J. Kim, K. Cai, J. Yoo, X. Fong, and H. Yang, Energy-efficient superparamagnetic Ising machine and its application to traveling salesman problems, Nature Communications, vol. 15, no. 3457, Apr 2024, doi: https://doi.org/10.1038/s41467-024-47818-z.

[16] M. A. Iftakher, H. Levices, K.-E. Harabi, A. Renaudineau, M.-C. Faye, C. Bouchard, F. Disdier, B. Viala, E. Vianello, P. Talatchian, K. Garello, D. Querlioz, and L. Hutin, Intrinsic Annealing in a Hybrid Memristor-Magnetic Tunnel Junction Ising Machine, arXiv:2506.14676, doi: https://doi.org/10.48550/arXiv.2506.14676.

[17] K. Y. Camsari, S. Salahuddin, and S. Datta, Implementing p-bits With Embedded MTJ, IEEE EDL, vol. 38, no. 12, pp. 1767-1770, Dec. 2017, doi: 10.1109/LED.2017.2768321.

[18] S. Jenkins, A. Meo, L. E. Elliott, S. K. Piotrowski, M. Bapna, R. W. Chantrell, S. A. Majetich, and R. H. L. Evans, Magnetic stray fields in nanoscale magnetic tunnel junctions, Jour. of Phys. D, vol. 53, pp. 044001, doi: 10.1088/1361-6463/ab4fbf.

**Figures**

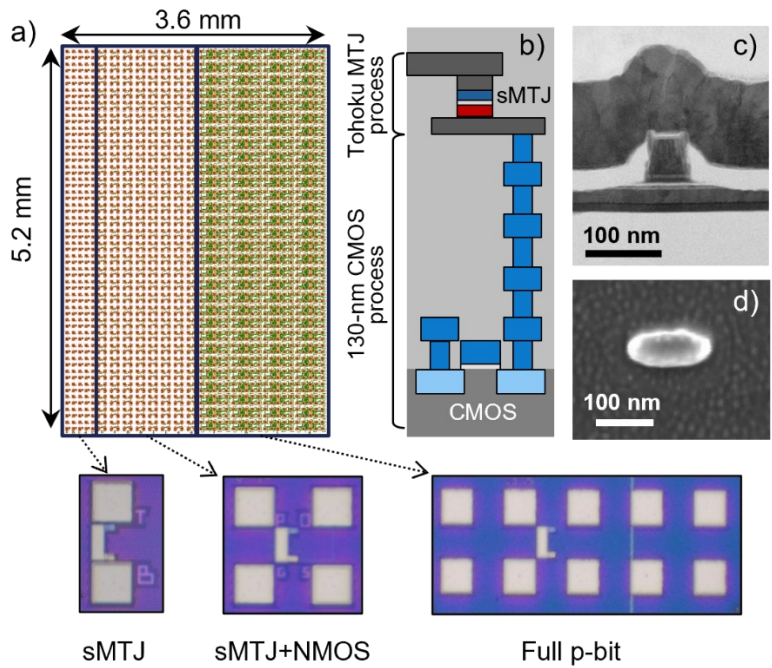

Fig. 1. a) Layout schematic of test chip with 150 isolated sMTJs, 240 sMTJs in series with an NMOS transistor, and 150 full-stage p-bit circuits. b) Schematic of the cross-sectional structure of the integrated chip. c) cross-sectional scanning transmission electron micrograph of an SMTJ integrated above the CMOS. d) Plan-view scanning electron micrograph of an sMTJ.

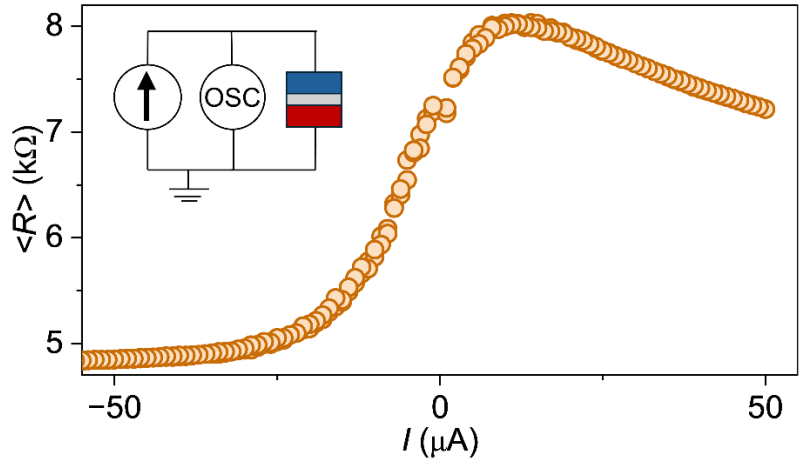

Fig. 2. Time-averaged resistance versus applied current for a single sMTJ with 25 nm × 100 nm dimensions. (inset) Experimental measurement setup.

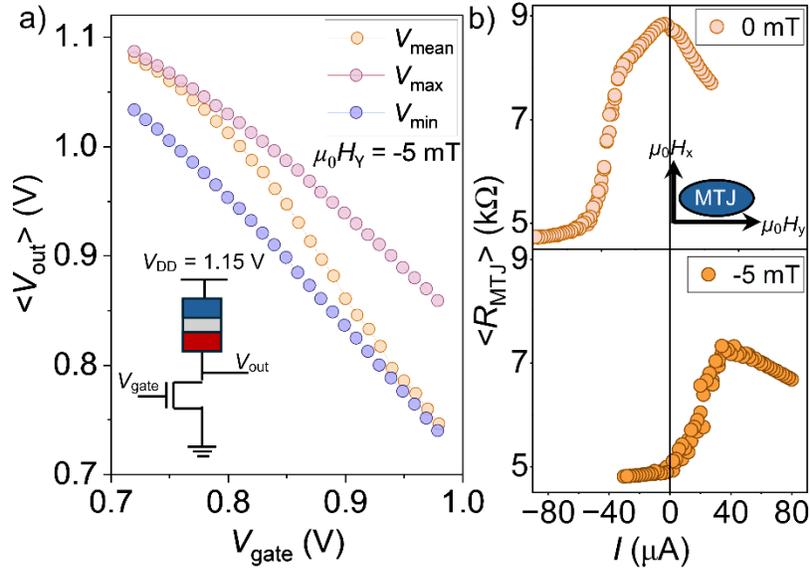

Fig. 3. a) Time-averaged characteristics of an sMTJ (dimensions: 25 nm × 100 nm) integrated in series with an NMOS transistor (channel width 1 μm). $V_{max}$ and $V_{min}$ represent the maximum and minimum voltages of $V_{out}$ read on the oscilloscope as $V_{gate}$ is increased. $V_{mean}$ represents the mean voltage measured, demonstrating the ability to tune the sMTJ state. (inset) Circuit schematic of an sMTJ integrated in series with an NMOS transistor. B) Time-averaged sMTJ resistance versus current plots showing the impact of applying an external magnetic field to shift the center of the sMTJ switching curve.

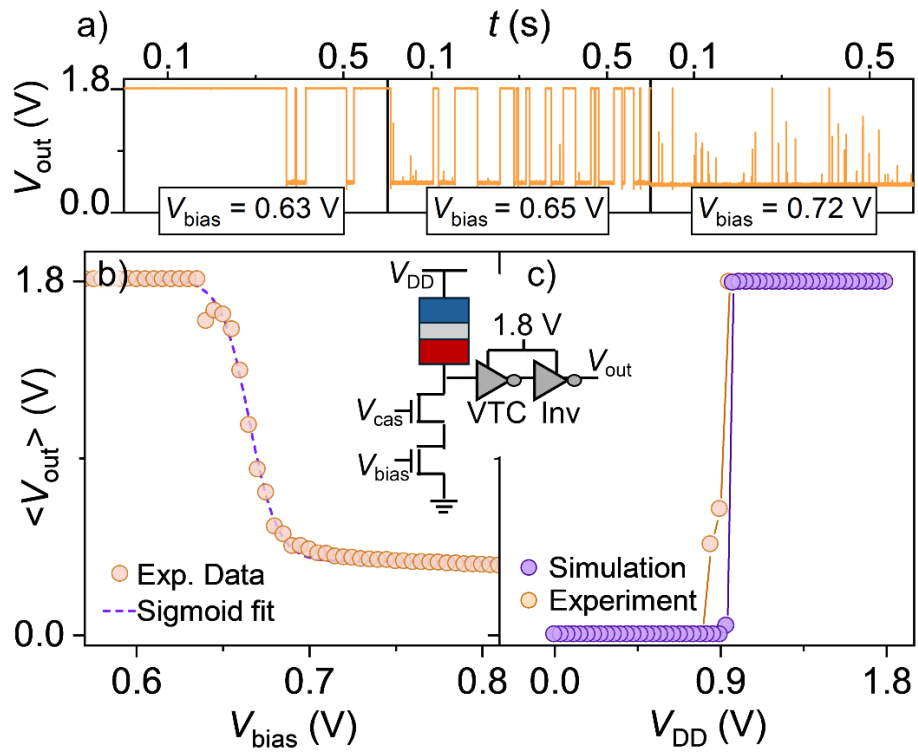

Fig. 4. a) Time-series measurements of $V_{out}$ for three $V_{bias}$ set points. The dimensions of the sMTJ for this device are 25 nm × 100 nm and the NMOS channel width is 1 μm. b) time-averaged characteristics of the same device shown in a). c) $V_{out}$ as a function of $V_{DD}$ when the p-bit input ($V_{cas}$, $V_{bias}$) is 0 V. (inset) Circuit schematic for the full-stage p-bit.